\documentclass{article}
\usepackage{graphicx} 
\usepackage{todonotes}
\usepackage{caption}
\usepackage{subcaption}
\usepackage{url}
\usepackage{hyperref}
\usepackage{orcidlink}
\usepackage{tabularx}
\usepackage{multirow}
\usepackage{hyperref}

\title{No Saved Kaleidosope: an 100\% Jitted Neural Network Coding Language with Pythonic Syntax}
\author{
Augusto Seben da Rosa\textsuperscript{1}\orcidlink{0009-0001-6773-2674},
Marlon Daniel Angeli\textsuperscript{1}, \\
Jorge Aikes Junior\textsuperscript{1}\orcidlink{0000-0001-6933-6168},
Alef Iury Ferreira\textsuperscript{3}\orcidlink{0000-0002-9119-6357}, \\
Lucas Rafael Gris\textsuperscript{3}, 
Anderson da Silva Soares\textsuperscript{3}, \\
Arnaldo Candido Junior\textsuperscript{4}\orcidlink{0000-0002-5647-0891}, 
Frederico Santos de Oliveira\textsuperscript{2}\orcidlink{0000-0002-5885-6747}, \\
Gabriel Trevisan Damke\textsuperscript{1}\orcidlink{0009-0001-9369-1722} and 
Rafael Teixeira Sousa\textsuperscript{2}\orcidlink{0000-0001-5998-046X} \\
 \\ \\
\textsuperscript{1} Federal University of Technology - Paraná \\
\textsuperscript{2} Federal University of Mato Grosso \\
\textsuperscript{3} Federal University of Goias \\
\textsuperscript{4} São Paulo State University \\
}
\date{}

\begin{document}

\maketitle

\begin{abstract}
We developed a jitted compiler for training Artificial Neural Networks using C++, LLVM and Cuda. It features object-oriented characteristics, strong typing, parallel workers for data pre-processing, pythonic syntax for expressions, PyTorch like model declaration and Automatic Differentiation. We implement the mechanisms of cache and pooling in order to manage VRAM, cuBLAS for high performance matrix multiplication and cuDNN for convolutional layers. Our experiments with Residual Convolutional Neural Networks on ImageNet, we reach similar speed but degraded performance. Also, the GRU network experiments show similar accuracy, but our compiler have degraded speed in that task. However, our compiler demonstrates promising results at the CIFAR-10 benchmark, in which we reach the same performance and about the same speed as PyTorch.
We make the code publicly available at: \href{https://github.com/NoSavedDATA/NoSavedKaleidoscope}{\url{https://github.com/NoSavedDATA/NoSavedKaleidoscope}}.
\end{abstract}

\section{Introduction}

Artificial Intelligence Neural Networks allowed the creation of models for label classification~\cite{resnet}, text generation~\cite{brown2020language, llama2}, image generation~\cite{stablediffusion}, robotics~\cite{rt-1, daydreamer} and beyond. The foundations of these models relies on coding languages such as Python and Julia\footnote{\href{https://julialang.org/}{\url{https://julialang.org/}}}, and on frameworks such as PyTorch~\cite{torch}, Tensorflow\footnote{\href{https://www.tensorflow.org/}{\url{https://www.tensorflow.org/}}} and Jax\footnote{\href{https://jax.readthedocs.io/en/latest/quickstart.html}{\url{https://jax.readthedocs.io/en/latest/quickstart.html}}}.
The success of these foundation is due to their simplicity, expressivity and model training speed.

In that regard, three main coding language types are considered when thinking about high performance computation: interpreted, compiled and jitted~\cite{jit}. The slowest one is the interpreted type -- like Python, it works by reading and compiling every line of code. This type of compiler adds computational overhead on the cases of for loops and function calls. Compiled languages will create an intermediate code file representation and so it does not need to recompile code. Jitted languages are on the mean term of both: it reads lines one by one like an interpreter, but the jitter organizes and saves the intermediate code of loops and functions~\cite{jit}. Thus, it avoids recompilation and it has a better speed than interpreted languages~\cite{jit_speed}.

Besides, Python parallelism is complex, it is not standardized across its libraries and it does not contain native support for GPU parallelism. For instance, in order to achieve true GPU parallelism, it would be required to utilize a library such as ray~\cite{efficientzero}.

At this paper, we describe an early version of the No Saved Kaleidoscope (NSK) coding language. Our contribution is as follows:  1. We demonstrate some of the processes required to create a language that is able to caught up with the Python-PyTorch model training speed with a Python like syntax 2. We harness the capabilities of LLVM to create a fully jitted language. 3. We implemented finish/async expressions as simple high-level parallelism code. 4. We present the Backpropagation~\cite{backpropagation} algorithm using a single line of code by using Automatic Differentiation~\cite{automatic_differentiation}.

%\arnaldo{a introdução está um pouco suscinta, é interessante dar mais contexto}

\section{Related Work}

Other neural network academic coding languages are RADENN~\cite{radenn}, which focuses on simplifying neural network and the data pipeline declaration; DeepScratch~\cite{deepscratch} that created a visual coding language with focus on facilitating the learning of deep learning; DeepDSL~\cite{deepdsl} which is a domain specific language embedded in Scala, and it was an early implementation of portable neural network code by compiling them into Java source code.

There are also other LLVM~\cite{llvm} based coding languages with the purpose of facilitating to write compute efficient mathematical code, like Julia. This language also supports neural networks training. Currently, there are 139 thousand github PyTorch repositories and 21 thousand Julia repositories\footnote{Access date: 2024-08-22}.

In terms of compiler related methods, we take fully inspiration from Python. It is an object-oriented, high-level language and that disposes the use of brackets in conditional and control structures by using indentation only. Its simplicity accelerates the development time when compared to other coding languages like C++ or Java~\cite{lang_productivity}. This characteristic made the coding language popular, being currently the most used coding language for training neural networks.

In terms of Python neural network frameworks, some of the most used are: TensorFlow, PyTorch and Jax. PyTorch and TensorFlow offer similar syntax, while Jax gives the coder fine grained control for jitting functions. Also, some of the Jax optimizations allowed a faster inference time of models\footnote{\href{https://github.com/openai/whisper/discussions/1277}{\url{https://github.com/openai/whisper/discussions/1277}}}. We take the design choices of PyTorch, while also jitting $100\%$ of the code.

% https://arxiv.org/pdf/1701.02284
% https://ieeexplore.ieee.org/abstract/document/10207040
% https://ieeexplore.ieee.org/stamp/stamp.jsp?arnumber=9439420
% https://arxiv.org/pdf/1701.02284

% https://scholar.google.com.br/scholar?hl=pt-BR&as_sdt=0%2C5&q=%22domain+specific%22+%22programming+language%22+%22deep+learning%22&btnG=

\section{Compiler Methods}

Kaleidoscope~\cite{kaleidoscope} is the basis of our compiler. It successfully generates code for condition and control structures and recursion calls. However, it does not contain support for multiple instructions inside these structures and it is not object-oriented. Also, its scope control is incompatible with multiple threads, since it creates only one scope per function, but we need to have multiple scopes for the same function if it is called by different threads.

We allow the use of multiple instructions by assigning a list of instructions for condition and control structures. This assignment is organized during the parsing phase~\cite{compilers}, and we parse it by indentation, like Python. Furthermore, in order to make the language object-oriented, we reverse engineer the Python self argument of class methods. That is, when an object calls a method we send the object name as the first argument of the function. Then, when there is a variable using the self expression, we concatenate the object name contained on the function first argument with the variable name.

Following this line of thought, we also generate a new scope at function calls and send it as the function second argument. Every variable that does not use the self expression is concatenated with the scope. We also pass the previous scope as the third argument, so we can send variables into the scope from which they were called when using return expressions. We hide these three arguments at the high-level language for simplicity and they are only present at user declared functions (non-native functions).

In order to implement parallelism, we followed the guide from the Bolt compiler~\cite{bolt}. That is, C++ has a function for thread that allows to create threads for functions by passing the pointer of these functions. Bolt uses LLVM to call this C++ thread function and it passes the pointers of its high level functions to create concurrency.

Also, we realized that everything in Kaleidoscope is executed as a function, and therefore we were able to send the function pointer of any code section. We implement parallelism using finish/async expressions, i.e, finish is a structure that contains a list of normal/serial instructions and async (parallel) instructions. It waits for the async and the serial instructions to finish before exiting the structure. Beyond that, we use scope control and add mutex locks at cpu variable attributions in order to avoid data races~\cite{data_race} -- otherwise they would lead to segmentation faults.

Besides, we had to use C++ char pointers instead of LLVM variable Allocas for representing strings, because with LLVM Allocas the code would crash when executing hundreds of thousands of function calls. It seems that LLVM was not properly managing memory on this situation, since our garbage collection of char pointers solved the issue.

\section{Neural Network Methods}

In this Section, we present the mechanisms of Automatic Differentiation, Memory Pooling and Operation Overlapping.

\subsection{Automatic Differentiation}

We leverage the pre-existing structures of our compiler to build the Backpropagation algorithm~\cite{backpropagation} using Automatic Differentiation~\cite{automatic_differentiation}.

Compilers are restricted by hardware to use the form of three address code during the intermediate-code generation phase~\cite{compilers} for numerical expressions. Also, coding languages traditionally organize parsed instructions into Abstract Syntax Trees (AST)~\cite{compilers}. Thus, it is straightforward to represent this expressions using a binary tree~\cite{introduction_to_algorithms}, like in Kaleidoscope.

\begin{figure}[htb]
    \centering
    \caption{AST and Backpropagation Tree}
    \label{fig:AST_Backprop}
    \begin{subfigure}[b]{0.25\textwidth}
        \includegraphics[width=\textwidth]{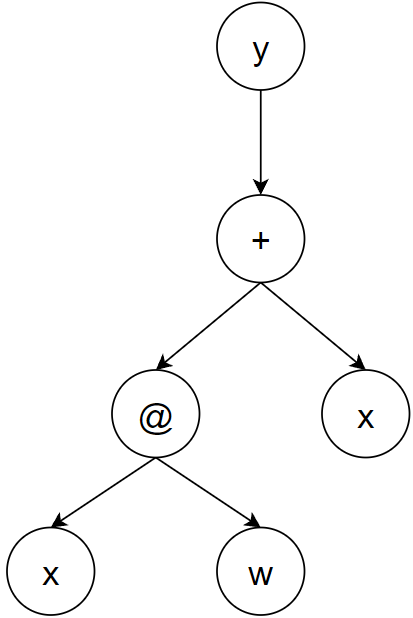}
        \caption{Mounting AST}
        \label{fig:ast}
    \end{subfigure}
    \begin{subfigure}[b]{0.3\textwidth}
         \includegraphics[width=\textwidth]{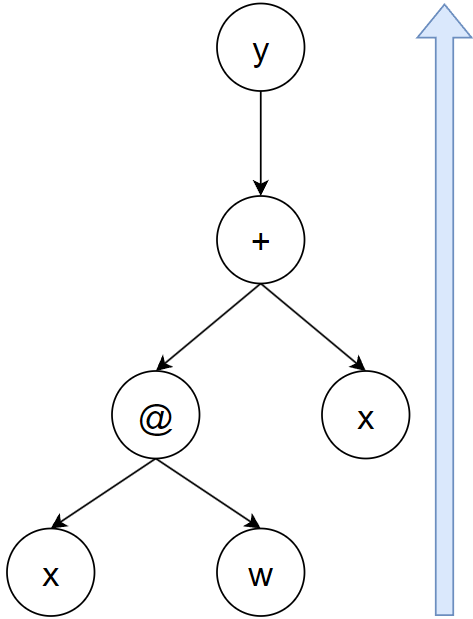}
         \caption{Forward}
         \label{fig:forward}
    \end{subfigure}
    \begin{subfigure}[b]{0.3\textwidth}
         \includegraphics[width=\textwidth]{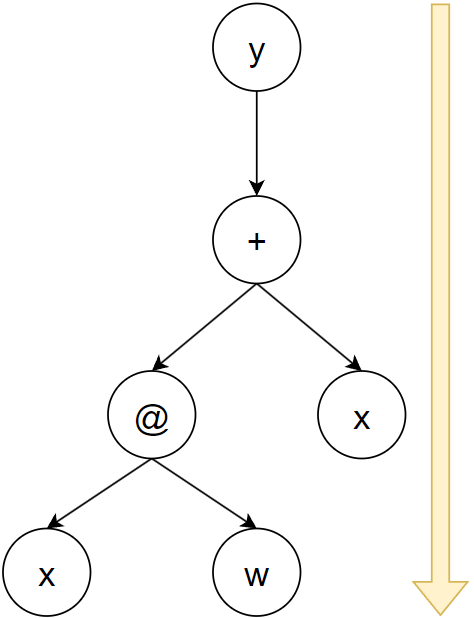}
         \caption{Backward}
         \label{fig:backward}
    \end{subfigure}
\end{figure}

This process is observed at Figure~\ref{fig:ast}, in which the parsing algorithm mounts a binary tree from the expression $y = x@w + x$, i.e, the assignment of $y$ from the result of adding $x$ with the matrix multiplication between $x$ and $w$, expressed by the operator $@$ -- which transposes $w$. The Kaleidoscope algorithm for this parsing organizes nodes according to their precedence: high precedence operators like $@$ should be solved first and thus they are close to leaf nodes, whereas low precedence operators like + and $=$ should be solved later.

After parsing and generating intermediate code for this binary tree and posterior ASTs, the expression code will be executed, Figure~\ref{fig:forward}. During this process, the instructions will travel from the leaves into the root (bottom-up). We save the results and operations of each node and recreate the binary tree. This way, we can traverse the tree on the opposite direction (top-down) for the backward mode.

Figure~\ref{fig:backward} illustrates the execution of the chain rule~\cite{backpropagation} of gradients. In other words, we load the gradient tensor stored at the variable $y$, then the same tensor gradient (float pointer) is sent to the left and right nodes of the $+$ operator\footnote{as this is the result of $+$ operation gradient}. On the right node, we accumulate the incoming gradient with the pre-existing gradient in $x$. And on the left node, the gradients with respect to $x$ and $w$ are each calculated according to the matrix multiply gradient rule (which we implement using cuBLAS\footnote{\url{https://developer.nvidia.com/cublas}}), and then their gradients are also accumulated.

With this in mind, after executing the backward binary tree of a single assignment (Figure~\ref{fig:backward}) it is also necessary to consider the case of multiple sequential assignments. This macro view of instruction shall also be executed on the reverse order of the forward mode. We therefore implement the macro structure as a stack of the aforementioned backward binary tree.

This way, the network first loads the data and the generated tensor node is also sent to the stack\footnote{Even though there is no gradient to be calculated during data loading or operations like the onehot, it is important to send these nodes for the Backpropagation algorithm because the same algorithm will execute the memory garbage collection.}, then the tensor will be passed to a neural network forward method, and each assignment operation will be represented as a binary tree that is pushed into the stack. The last node on the stack must be from a loss function\footnote{This has the advantage of forcing the coder to use loss optimized kernels.}, and after this we begin to pop from the stack. Each pop will load the accumulated gradient from the assigned tensor and then execute the already mentioned algorithm for that backward binary tree. With this, we pop from the stack and repeat this process until it is empty.

%\arnaldo{citar mesmo que brevemente a regra da cadeia para amarrar com diferenciação automática e ast}

\subsection{Caching and Pooling}

Beyond the Backpropagation, it is also important to apply efficient memory management algorithms. PyTorch applies the mechanisms of Caching and Pooling to maximize this efficiency, and we derive from these mechanisms for NSK.

We implement caching for memory information that should be kept across multiple forward and backward iterations. For example, the memory of weight and biases gradients: they consist of a global variable dictionary that maps a parameter's name to its float pointer. This pointer is accumulated every time it is seen during the Backpropagation and it is overwritten with zeros after the optimizer finishes its iteration.

\begin{figure}[htb]
    \centering
    \caption{Naive Memory Management}
    
    \includegraphics[width=0.65\textwidth]{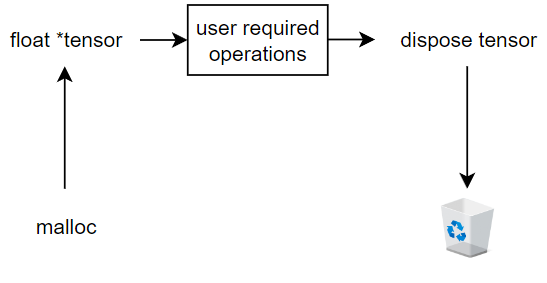}
    %\caption{Mounting AST}
    \label{fig:naive_memory}
\end{figure}

Furthermore, the naive approach to deal with tensors that are allocated at each iteration -- intermediate operation results and non-weight gradients -- is to malloc and free their pointers at every iteration, as illustrated in Figure~\ref{fig:naive_memory}. However, when tensors (gpu float pointers) have the size of millions of floats, this repetitive operation adds a heavy computational overhead.

% continuar leitura daqui
Hence, memory pooling is a technique applied to increase the efficiency of frequent allocation and deallocation of memory, like on the Linux Kernel~\cite{pooling_linux}. Besides, PyTorch also uses memory pooling for the memory management of tensors. Thus, we also apply memory pooling for tensors that should be allocated and deallocated at each iteration.

\begin{figure}[htb]
    \centering
    \caption{A User Samples a 512 Tensor from the Pool}
    
    \includegraphics[width=\textwidth]{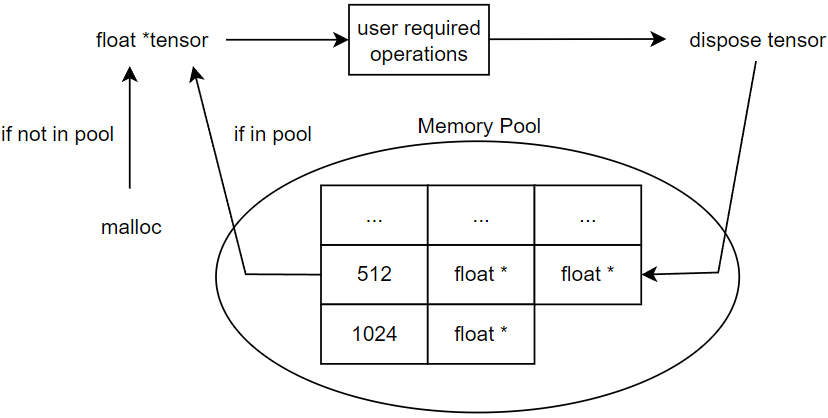}
    \label{fig:memory_pool}
\end{figure}

As demonstrated by Figure~\ref{fig:memory_pool}, we create a dictionary that maps the size of a required tensor into a list of gpu float pointers containing that size. The list of pointers for a given size will be empty the first time a tensor is required, we thus malloc a new pointer with that demanded size on the gpu on this case and send it to the operation it was required from. When the tensor is no longer needed, we append its pointer to the list of pointers containing that size on the dictionary instead of freeing it. Now that the pool is not empty anymore, we sample a float pointer from the pool whenever required instead of allocating it.

The pool will be empty on the first iteration and it will be populated with all the required pointers on the second iteration onwards. That is, not a single tensor will be allocated by malloc after the first iteration.

One of our preliminary ResNet~\cite{resnet} CIFAR-10~\cite{cifar} experiments demonstrated that memory pooling improved the training speed beyond 5 times faster.

\subsection{Operation Overlapping}

Besides low-level Cuda kernel optimizations -- like kernel fusing~\cite{optimizing_lstm, flashattention}, operation overlapping is the last PyTorch optimization we are aware of.

To design Cuda efficient kernels it is desired to maximize GPU occupancy~\cite{optimizing_lstm, flashattention2}. In other words, when we launch kernels that are cheap to compute, we want to launch as many parallel kernels as possible. One of the ways PyTorch does this is by overlapping data loading and model training using different Cuda streams. This means that while a network is processing its batch, PyTorch will already start to load the next batch on the GPU because these operations belong to different streams.

We apply this overlapping technique at our compiler. We also launch a separate stream for each parameter group of the optimizer.

\section{Experiments and Discussion}

The Python experiments were run using Python 3.10.14, PyTorch 2.2.2 and cuDNN\footnote{\url{https://developer.nvidia.com/cudnn}} 8.9.2. NSK used cuDNN 8.9.7. We used Cuda 12.1 for both PyTorch and NSK. The hardware comprises a RTX 4090, a i9-13900KF and 32 GB of RAM.

We execute CNN experiments on MNIST~\cite{mnist}, CIFAR-10~\cite{cifar} and ImageNet~\cite{imagenet}. And the Recurrent Neural Networks (RNN) experiments at the IDMB Sentiment Analysis benchmark~\cite{imdb_sentiment_analysis}. We use 3 parallel workers for all experiments except ImageNet, in which we use 12. Also, we implemented the image augmentations at the GPU instead of at the CPU workers.

Currently, NSK does not support dictionaries. Thus, we had to change the folder name of each dataset class to a class number. Then, we splitted the file path string and associated the instance class to that number.

We follow the ResNet recipe of~\cite{resnet} with 20 layers for the CIFAR-10 benchmark. The only differences we are aware of is the weight initialization -- we use xavier uniform~\cite{zhang2023dive}, and the data augmentations: we use random crop with padding 4 and random horizontal flip. This benchmark results are demonstrated in Table~\ref{tab:cifar}.

The column ``DS From Scratch'' involves training the network while implementing the dataset on PyTorch from scratch. The ``CIFAR-10'' column uses the PyTorch CIFAR10 class. This version downloads the dataset into disk, but it saves and reads the data as numpy tensors instead of images. The ``NSK column'' also loads images from disk. We run these experiments for 10 seeds. The ``Time row'' is the time avarage across the last 5 seeds. The VRAM usage had a standard deviation below 100 MB across all runs.

At the CIFAR-10 benchmark, we obtained poor results with the shuffle option of the builtin torch Dataloader class combined with our Dataset implementation from scratch. For instance, we achieved a mean accuracy of $17.83\pm0.26$ for 3 seeds. Thus, we shuffle the file list using the random library instead.

\begin{table}[htb]
\centering
\caption{CIFAR-10 using a 20-layers ResNet}
\label{tab:cifar}
%\resizebox{\textwidth}{!}{
\begin{tabular}{l r r r r} 
 \hline
 \textbf{Stat} & \textbf{DS From Scratch} & \textbf{CIFAR-10} & \textbf{NSK} \\ [0.5ex] 
 \hline
 Accuracy  & 87.46$\pm$1.06 & 87.2$\pm$0.45 & 87.80$\pm$0.59  \\ 
 Training VRAM & 2.2 GB & 2.2 GB & 2.8 GB \\
 Time & 9m 24s $\pm$ 3s & 6m 29s $\pm$ 4s & 7 m 43s $\pm$ 7s \\
 \hline
\end{tabular}
%}
\end{table}

To obtain the CIFAR-10 image file version, we first run the PyTorch CIFAR10 class to download the images. Then, we read each numpy tensor and save it as an image using a PyTorch transform.

We have also experimented testing cuDNN 9 for PyTorch with the dataset implemented from scratch, but NSK was still around 20 seconds faster that PyTorch on that case. Besides, it is fairer to compare NSK to the ``DS From Scratch'' column, since they are both loading images from disk.

Additionally, we follow the ResNet 18 recipe \cite{resnet} for the ImageNet benchmark. However, we change the learning rate scheduler by the cosine scheduler \cite{cosine_lr} and lower the amount of training steps to 10.000 and 30.000 instead of 600.000. Also, we kept the data augmentations used at the CIFAR-10 benchmark. The results are presented at Table \ref{tab:imagenet}. We train the networks for a single seed. Here, Torch From Scratch representing implementing the GRU from scratch instead of using the nn.GRU class, and the Embedding was substituted by a one.

\begin{table}[htb]
\centering
\caption{ImageNet with ResNet 18}
\label{tab:imagenet}

\begin{tabular}{l r r r r} 
 \hline
 \multicolumn{2}{c}{\textbf{Stat}} & \textbf{DS From Scratch} & \textbf{NSK} \\ [0.5ex] 
 \hline
    \multirow{2}{*}{10k steps} & Acc & \multicolumn{1}{r}{45.24\%} & \multicolumn{1}{r}{39.29\%} \\
            & Time & \multicolumn{1}{r}{56m} & \multicolumn{1}{r}{1H 35m} \\\cline{3-4}
            
    \multirow{2}{*}{30k steps} & Acc & \multicolumn{1}{r}{59.97} & \multicolumn{1}{r}{46.1\%} \\
            & Time & \multicolumn{1}{r}{3H 45m} & \multicolumn{1}{r}{3H 30m} \\\cline{3-4}
 \multicolumn{2}{l}{Training VRAM} & 12 GB & 24 GB \\
 
 \hline
\end{tabular}
\end{table}

We failed to implement the ImageNet networks, as demonstrated by Table~\ref{tab:imagenet}. We believe this is due to an uncaught out of memory error of the compiler, because memory allocation for the evaluation phase is severally slower than the training phase.

Besides, we test the GRU RNN~\cite{gru} at the IMDB Sentiment Analysis benchmark. The network comprises an embedding layer with the size of the vocabulary, one GRU layer with hidden size 256 and the output layer. We train it for 4 epochs with a batch size of 50, sequence length 200, vocabulary size of 32768 and using cross entropy with 2 classes. We used the AdamW~\cite{adamw} optimizer with a fixed learning rate of 0.001, weight decay of 0.0001 and gradient clip of 5. Its results are presented at Table~\ref{tab:gru}, using 3 seeds.

\begin{table}[htb]
\centering
\caption{IMDB Sentiment Analysis using a GRU Network}
\label{tab:gru}
\begin{tabular}{l r r r} 
 \hline
 \textbf{Stat} & \textbf{Torch From Scratch} & \textbf{Torch} & \textbf{NSK} \\ [0.5ex] 
 \hline
 Accuracy  & 80.63$\pm$0.06 & 82.72$\pm$0.39 & 83.07$\pm$0.48  \\ 
 Training VRAM & 8 GB & 8 GB & 12 GB \\
 Inference VRAM & 13 GB & 13 GB & 15 GB \\
 Time & 207s $\pm$ 4s & 42s $\pm$ 2s & 647s $\pm$ 187s \\
 \hline
\end{tabular}
\end{table}

PyTorch presents a 20 times faster GRU network when compared to NSK. Since our cuDNN implementation matches their CNN speed, we believe they are also using the cuDNN versions of RNNs~\cite{optimizing_lstm}, whereas we are using cuBLAS without any kernel fusion optimizations.

Overall, PyTorch demonstrates a memory pooling mechanism that is slightly to two times more efficient than NSK.

\section{Conclusion}

We have implemented a jitted coding language that supports concurrency, object orientation and neural network training. Despite our Backpropagation implementation successfully trained residual and recurrent neural networks, there is still need of a better memory pooling, speed improvements and higher architectures support.

For future work, we need to improve the efficiency of the memory pooling mechanism. Furthermore, we noticed that on the PyTorch repository there are several operations which have a fused version with the Relu~\cite{relu} activation function. We pretend to apply this type of operation fusion and to optimize recurrent neural networks as well, similar to~\cite{optimizing_lstm}.

On the long term, we pretend to add support of Generative Adversarial Networks~\cite{gan_goodfellow}, Transformers~\cite{vaswani2017attention}, Diffusion Models~\cite{stablediffusion}, Reinforcement Learning and Audio Neural Networks. We will also implement class inheritance on the future.

%\augusto{Precisa dessas citações aqui?}

\section{Acknowledgements}

This work has been fully/partially funded by the project supported by Acknowledge Center of Immersive Technologies (AKCIT), with financial resources from the PPI IoT/Manufatura 4.0 / PPI HardwareBR of the MCTI grant number 057/2023, signed with EMBRAPII.

Also, we wish to thank for the intellectual support of Karpathy for organizing multiple complex Cuda/C++ codes for training a GPT model into a single github repository, the Kaleidoscope coding language tutorial for the basis of our compiler and the Bolt compiler documentation that allowed us to implement concurrency and therefore the parallel workers.

Without any of these contributions, our project would have been impossible.

\bibliographystyle{abbrv} % ver abaixo para opções
\bibliography{main} % incluir referencias.bib

\end{document}